\documentclass[fleqn,10pt]{article}
\usepackage{epsfig}
\newcommand{\be}{\begin{equation}}
\newcommand{\ee}{\end{equation}}

\newcommand{\eq}[1]{eq.~(\ref{#1})}
\newcommand{\fig}[1]{Fig.~\ref{#1}}
\newcommand{\chii}{\chi_{{}_{\rm I}}}
\newcommand{\chir}{\chi_{{}_{\rm R}}}
\newcommand{\pbar}{\bar p}

\def\spai{\sigma_{p{-}\rm air}^{\rm inel}}

\def\signn{\sigma_{\rm nn}}
\def\siggp{\sigma_{\gamma\rm p}}
\def\siggg{\sigma_{\gamma\gamma}}
\def\rhonn{\rho_{\rm nn}}
\def\rhogp{\rho_{\gamma\rm p}}
\def\rhogg{\rho_{\gamma\gamma}}
\def\Bnn{B_{\rm nn}}
\def\Bgp{B_{\gamma\rm p}}
\def\Bgg{B_{\gamma\gamma}}
\begin{document}
\setcounter{secnumdepth}{4}
\renewcommand\thepage{\ }
%
%
\begin{titlepage} 
%
\newcommand\reportnumber{901} 
\newcommand\mydate{November 26, 2001} 
\newlength{\nulogo} 
\settowidth{\nulogo}{\small\sf{N.U.H.E.P. Report No. \reportnumber}}
\title{
\vspace{-.8in} 
\hfill\fbox{{\parbox{\nulogo}{\small\sf{Northwestern University: \\
N.U.H.E.P. Report No. \reportnumber\\
University of Wisconsin:\\
MADPH-01-1251\\
revised:\ 
          \mydate}}}}
\vspace{0.5in} \\
{
On factorization, quark counting and vector dominance
}}

\author{
M.~M.~Block
\thanks{Work partially supported by Department of Energy contract
DA-AC02-76-Er02289 Task D.}\vspace{-5pt}   \\
{\small\em Department of Physics and Astronomy,} \vspace{-5pt} \\ 
{\small\em Northwestern University, Evanston, IL 60208}\\
\vspace{-5pt}
\  \\
F.~Halzen
\thanks{Work partially supported by Department of Energy 
Grant No.~DE-FG02-95ER40896 and the University of Wisconsin Research
Committee with funds granted by the Wisconsin Alumni Research
Foundation.}
\vspace{-5pt} \\ 
{\small\em Department of Physics,} 
\vspace{-5pt} \\ 
{\small\em University of
Wisconsin, Madison, WI 53706} \\
\vspace{-5pt}\\
%
%
%
%
G.~Pancheri
\vspace{-5pt} \\
{\small\em INFN-Laboratori Nazionali di Frascati,}\vspace{-5pt}  \\
{\small\em Frascati, Italy}\\
\vspace{-5pt}\\
}    
\vspace{.5in}
\date {}
\maketitle

\begin{abstract}
Using an eikonal structure for the scattering amplitude, Block and Kaidalov\cite{BK} have derived factorization theorems for nucleon-nucleon, $\gamma p$ and $\gamma\gamma$ scattering at high energies, using only some very general assumptions.
We present here an analysis giving experimental confirmation for factorization of cross sections, nuclear slope parameters B and $\rho$-values (ratio of real to imaginary portion of forward scattering amplitudes), showing that:
\begin{itemize}
\item the three factorization theorems\cite{BK} hold,
\item the additive quark model holds to $\approx 1\%$,
\item and vector dominance holds to better than $\approx 4\%$.
\end{itemize}  
\end{abstract}
\thispagestyle{empty}
\end{titlepage}
\newpage
\pagenumbering{arabic}
\renewcommand{\thepage}{-- \arabic{page}\ --} 
\section{Introduction}
Assuming  factorizable eikonals in impact parameter space $b$ for nucleon-nucleon, $\gamma$p and $\gamma\gamma$ scattering processes whose opacities are equal, Block and Kaidalov\cite{BK} have proved three  factorization theorems:
\begin{enumerate}
\item \label{enum:sig}  \[ \frac{\signn(s)}{\siggp(s)}=\frac{\siggp(s)}{\siggg(s)},\] where the $\sigma$'s are the total cross sections for nucleon-nucleon, $\gamma$p and $\gamma\gamma$ scattering,
\item \label{enum:B} \[ \frac{\Bnn(s)}{\Bgp(s)}=\frac{\Bgp(s)}{\Bgg(s)},\] where the $B$'s are the nuclear slope parameters for elastic scattering,  
\item \label{enum:rho}\[ \frac{\rhonn(s)}{\rhogp(s)}=\frac{\rhogp(s)}{\rhogg(s)},\] where the $\rho$'s are the ratio of the real to imaginary portions of the forward scattering amplitudes,  
\end{enumerate}
with each factorization theorem having its own proportionality constant.  These theorems are exact, for {\em all } $s$  (where $\sqrt s$ is the c.m.s. energy), and survive exponentiation of the eikonal\cite{BK}.

Physically, the assumption of equal opacities, where the opacity is defined as the value of the eikonal at $b=0$, is the same as demanding that the ratios of elastic to total cross sections are equal, {\em i.e.,}
\be
\left(\frac{\sigma_{\rm el}}{\sigma_{\rm tot}}\right)_{\rm nn}= \left(\frac{\sigma_{\rm el}}{\sigma_{\rm tot}}\right)_{\gamma\rm p}=\left(\frac{\sigma_{\rm el}}{\sigma_{\rm tot}}\right)_{\gamma\gamma},\label{eq:sigeloversigtot}
\ee
as the energy goes to infinity\cite{BK}.

Factorization theorem \ref{enum:sig}, involving ratios of cross sections, is perhaps the best known.  Factorization theorems \ref{enum:B} and \ref{enum:rho} are less known, but turn out to be of primary importance.  The purpose of this note is to present strong experimental evidence for all three factorization theorems, as well as evidence for the additive quark model and vector dominance.
\section{Eikonal Model}
In an eikonal model\cite{blockhalzenpancheri}, a (complex)  
eikonal $\chi(b,s)$ is defined such that $a(b,s)$, the (complex) scattering
amplitude in impact parameter space $b$, is given by
\be
a(b,s)=\frac{i}{2}\left(1-e^{i\chi(b,s)}\right)
=\frac{i}{2}\left(1-e^{-\chii(b,s)+i\chir(b,s)}
\right).\label{eik}
\ee
Using the optical theorem,
the total cross section $\sigma_{\rm tot}(s)$ is given by
\begin{equation}
\sigma_{\rm tot}(s)=2\int\,\left[1-e^{-\chii
(b,s)}\cos(\chir(b,s))\right]\,d^2\vec{b},\label{sigtot}
\end{equation}
the elastic scattering cross section
$\sigma_{\rm el}(s)$ is given by
\begin{eqnarray}
\sigma_{\rm
elastic}(s)&=&\int\left|1-e^{-\chii(b,s)+i\chir(b,s)}\right|^2\,d^2\vec{b}\label{sigel}
\end{eqnarray}
and the inelastic cross section, $\sigma_{\rm inelastic}(s)$, is given by
\be
\sigma_{\rm inelastic}(s)=\sigma_{\rm tot}(s)-\sigma_{\rm
elastic}(s)=\int\,\left [1-e^{-2\chii(b,s)}\right
]\,d^2\vec{b}.\label{sigin}
\ee
The ratio of the real to the imaginary part of the forward nuclear
scattering amplitude, $\rho$,
is given by
\begin{eqnarray}
\rho(s)&=&\frac{{\rm Re}\left\{i(\int
1-e^{-\chii(b,s)+i\chir(b,s)})\,d^2\vec{b}\right\}}
{{\rm Im}\left\{i(\int
(1-e^{-\chii(b,s)+i\chir(b,s)})\,d^2\vec{b}\right\}}\label{rho}
\end{eqnarray}
and the nuclear slope parameter $B$ is given by
\be
B=
\frac{\int\,b^2a(b,s)\,d^2\vec{b}}{2\int\,a(b,s)\,d^2\vec{b}}.\label{Bsimple}
\ee
\subsection{Even Eikonal\label{app:QCDeven}}
A description of the  
forward proton--proton and proton--antiproton scattering amplitudes is  
required which is analytic, unitary, satisfies crossing symmetry and the Froissart bound. A  
convenient parameterization\cite{block,blockhalzenpancheri} consistent with the above constraints and with the high-energy data can be  
constructed in a model where the asymptotic nucleon becomes a black  
disk as a reflection of particle (jet) production. The increase of the  
total cross section is the shadow of jet-production which is  
parameterized in parton language. The picture does not reproduce the  
lower energy data which is simply parameterized using Regge  
phenomenology. The even QCD-inspired eikonal $\chi_{\rm even}$ for nucleon-nucleon scattering\cite{block,blockhalzenpancheri} is  
given by the sum of
three contributions, gluon-gluon, quark-gluon and quark-quark, which
are individually factorizable into a product of a cross section
$\sigma (s)$ times an impact parameter space distribution function
$W(b\,;\mu)$,  {\em i.e.,}:
\begin{eqnarray}
 \chi^{\rm even}(s,b)& = &\chi_{\rm gg}(s,b)+\chi_{\rm
qg}(s,b)+\chi_{\rm qq}(s,b)\nonumber\\
&=&i\left[\sigma_{\rm gg}(s)W(b\,;\mu_{\rm gg})+\sigma_{\rm
qg}(s)W(b\,;\sqrt{\mu_{\rm qq}\mu_{\rm gg}})+\sigma_{\rm
qq}(s)W(b\,;\mu_{\rm qq})\right],\label{eq:chieven}
\end{eqnarray}
where the impact parameter space distribution function is the
convolution of a pair of dipole form factors:
\begin{equation}
W(b\,;\mu)=\frac{\mu^2}{96\pi}(\mu b)^3K_3(\mu b).\label{W}
\end{equation}
It is normalized so that $\int W(b\,;\mu)d^2 \vec{b}=1.$ Hence, the
$\sigma$'s in \eq{eq:chieven} have the dimensions of a cross section.  
The factor $i$ is inserted in \eq{eq:chieven} since the high energy
eikonal is largely imaginary (the $\rho$ value for nucleon-nucleon
scattering is rather small).

The opacity of the eikonal, its value at $b=0$, is given by 
\be
{\rm O}^{\rm nn}=\frac{i}{12\pi}\left[\sigma_{\rm gg}(s)\mu_{\rm gg}^2+\sigma_{\rm qg}(s)\mu_{\rm qg}^2
+\sigma_{\rm qq}(s)\mu_{\rm qq}^2\right],\label{opacitynn}
\ee
 a simple sum of the products of the appropriate cross sections $\sigma$ with the $\mu^2$'s, a result which we will utilize later.
\subsection{Odd Eikonal}
The odd eikonal, $\chi_{odd}(b,s)=i\sigma_{odd}\,W(b;\mu_{odd})$,
accounts for the difference between $pp$ and $p \bar{p}$, and must
vanish at high energies. A Regge behaved analytic odd eikonal can be
parametrized as (see Eq.~(5.5b) of Ref.~\cite{bc})
\begin{equation}
\chi_{odd}(b,s)
= -\,C_{odd}\Sigma_{gg}\frac{m_0}{\sqrt{s}}
\,e^{i\pi/4}\,W(b;\mu_{odd}) \,
\label{oddanalytic}
\end{equation}
where $\mu_{\rm odd}$ is determined by experiment and the normalization constant $C_{odd}$ is to be fitted. 
\subsection{Total Eikonal}
The data for both $pp$ and $\pbar p$ are fitted using the total eikonal
\be
\chi^{\pbar p}_{pp}=\chi_{\rm even}\pm \chi_{\rm odd}.\label{eq:totalchi}
\ee
\section{A Global Fit of Accelerator and Cosmic Ray Data}

\begin{figure}[h,t,b] 
\begin{center}
\mbox{\epsfig{file=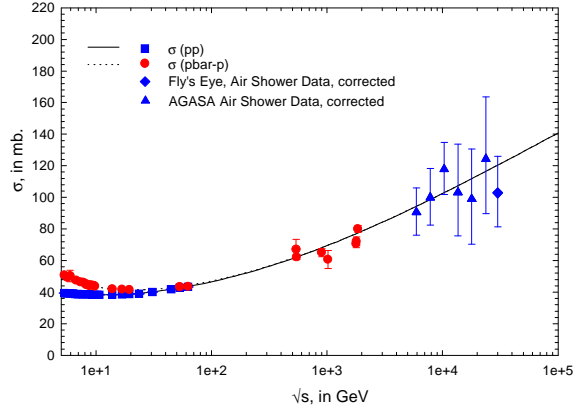%
,
width=3.4in,bbllx=75pt,bblly=370pt,bburx=525pt,bbury=675pt,clip=%
}}
\end{center}
\caption[] {\footnotesize
 The fitted $\sigma_{pp}$ and $\sigma_{\bar pp}$, in mb {\em vs.}
$\sqrt s$, in GeV, for the QCD-inspired fit of  total cross section,
$B$ and $\rho$  for both $pp$ and $\bar pp$. The accelerator data
(squares are pp and circles are $\bar {\rm   p}$p ) and  the cosmic ray  
points (diamond, Fly's Eye and triangles, AGASA) have been fitted
simultaneously. The cosmic ray data that are shown have been converted  
from  $\spai$ to $\sigma_{pp}$.}
\label{fig:ppcurves}
\end{figure}

\begin{figure}[h] 
\begin{center}
\mbox{\epsfig{file=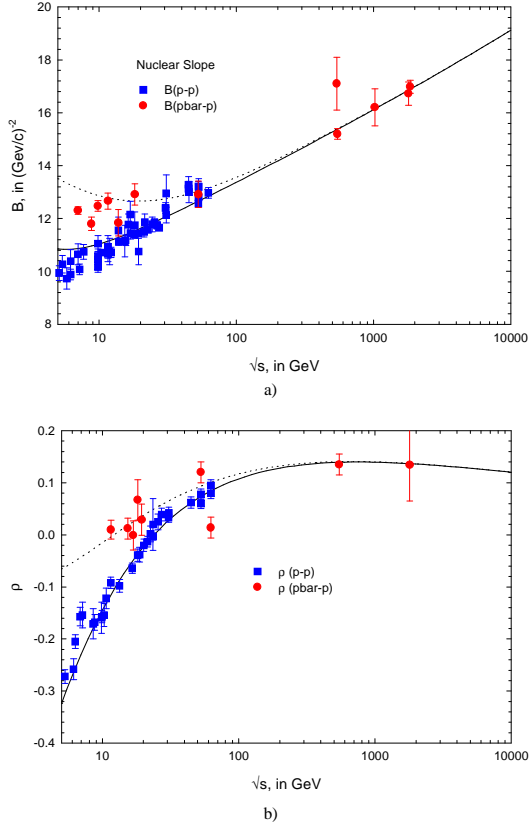%
            ,width=3.4in,bbllx=56pt,bblly=119pt,bburx=512pt,bbury=730pt,clip=%
}}
\end{center}
\caption[] {\footnotesize
 The fitted values for the nuclear slope parameters $B_{pp}$ and
$B_{\bar p p}$,  in (GeV/c)$^{-2}$ {\em vs.} $\sqrt s$, in GeV, for the  
QCD-inspired fit are shown in (a). In (b), the fitted values for
$\rho_{\bar pp}$ and $\rho_{pp}$ are shown.}
\label{fig:bandrho}
\end{figure}

Using an eikonal analysis in impact parameter space, Block {\em et al.}\,\cite{block,blockhalzenpancheri,blockcr} have constructed a QCD-inspired parameterization of the forward
 proton--proton and proton--antiproton scattering  
amplitudes which fits  all accelerator data\cite{orear}  
for
$\sigma_{\rm tot}$, nuclear slope parameter $B$
 and $\rho$, the ratio of the real-to-imaginary part of the forward
 scattering amplitude for both $pp$ and $\pbar p$ collisions, using a $\chi^2$ procedure and the eikonal of \eq{eq:totalchi}; see
Fig.\,\ref{fig:ppcurves} and Fig.\,\ref{fig:bandrho} 
%
which are taken from ref. \cite{blockcr}---in addition, the  
high energy cosmic ray cross sections of Fly's Eye\,\cite{fly} and
AGASA\,\cite{akeno} experiments are also {\em simultaneously}
fit\cite{blockcr}.  %
Because the  parameterization is
 both unitary and analytic, its high energy predictions are
 effectively model--independent, if you require that the proton is
asymptotically a black disk.  A 
major difference between  simultaneously  fitting the cosmic ray and accelerator  
data
and earlier results in which only accelerator data were 
used, is a lowering (by about a factor of $\approx 2$) of the  
error  of the predictions for the high energy cross sections. In particular, the error in $\sigma_{pp}$ at $\sqrt s=30 $ TeV is reduced to $\approx 1.5$ \%, because of significant reductions in the errors estimated for the fit parameters (for a more complete explanation, see ref. \cite{blockcr}).

The plot of $\sigma_{pp}$ {\em vs.} $\sqrt s$, including the  cosmic
ray data,  is
shown in Fig.\,\ref{fig:ppcurves}, which was taken from ref. \cite{blockcr}. The overall agreement between the accelerator and the cosmic
ray $pp$ cross sections with the QCD-inspired fit, as shown in
Fig.\,\ref{fig:ppcurves}, is striking.  

In brief, the eikonal description provides an excellent description of the experimental data at high energy for both $pp$ and $\pbar p$ scattering at high energies.  
%
%
\section{Factorization}
We emphasize that the QCD-inspired  parameterization of the $pp$ and $\pbar p$ data\,\cite{block,blockhalzenpancheri,blockcr}  allows us to calculate accurately the {\em even} eikonal of \eq{eq:chieven} needed for:
\begin{itemize}
\item the total cross section $\signn$ (from \eq{sigtot})  used in the factorization theorem \ref{enum:sig}, 
\item the nuclear slope parameter $\Bnn$ (from \eq{Bsimple}) used in the factorization theorem \ref{enum:B},
\item and the $\rho$-value  $\rhonn$ (from \eq{rho}) used in the factorization theorem \ref{enum:rho},
\end{itemize}
since we must compare nn to $\gamma p$ and $\gamma\gamma$ reactions.
%
%
\subsection{Theorems} 
As shown in ref. \cite{BK}, the eikonals for $\gamma p$ and $\gamma \gamma$ scattering that satisfy \eq{eq:sigeloversigtot}are given by
\be
\chi^{\gamma p}(s,b)= i\left[\kappa\sigma_{\rm
gg}(s)W(b\,;\sqrt{\frac{1}{\kappa}}\mu_{\rm gg})+\kappa\sigma_{\rm
qg}(s)W(b\,;\sqrt{\frac{1}{\kappa}}\sqrt{\mu_{\rm qq}\mu_{\rm
gg}}+\kappa\sigma_{\rm qq}(s)W(b\,;\sqrt{\frac{1}{\kappa}}\mu_{\rm
qq})\right], \label{eq:chigp}
\ee
and
\be
\chi^{\gamma \gamma}(s,b)= i\left[\kappa^2\sigma_{\rm
gg}(s)W(b\,;\frac{1}{\kappa}\mu_{\rm gg})+\kappa^2\sigma_{\rm
qg}(s)W(b\,;\frac{1}{\kappa}\sqrt{\mu_{\rm qq}\mu_{\rm
gg}}+\kappa^2\sigma_{\rm qq}(s)W(b\,;\frac{1}{\kappa}\mu_{\rm
qq})\right], \label{eq:chigg}
\ee
where we obtain $\chi^{\gamma p}$ from $\chi^{\rm even}$ from multiplying each $\sigma$ in $\chi^{\rm even}$ by $\kappa$ and each $\mu$ by $\sqrt\frac{1}{\kappa}$, and, in turn, we next obtain $\chi^{\gamma \gamma}$ from $\chi^{\gamma p}$ from multiplying each $\sigma$ in $\chi^{\gamma p}$ by $\kappa$ and each $\mu$ by $\sqrt\frac{1}{\kappa}$. The $\kappa$ in \eq{eq:chigp} and \eq{eq:chigg} is an energy-independent proportionality constant. We emphasize that the {\em same} $\kappa$ must be used in the gluon sector as in the quark sector for the ratio of $\left(\frac{\sigma_{\rm el}}{\sigma_{\rm tot}}\right)$ to be process-independent (see \eq{eq:sigeloversigtot}.  The functional forms of the impact parameter distributions are assumed to be the same for $\gamma p$, $\gamma \gamma$ and $nn$ reactions. It is clear from using \eq{W}, \eq{eq:chigp}, \eq{eq:chigg} and then comparing to the opacity of \eq{opacitynn}, that the three opacities are all the same, {\em i.e.,}
\be
O^{\rm nn}=O^{\gamma p}=O^{\gamma\gamma}=\frac{i}{12\pi}\left[\sigma_{\rm gg}(s)\mu_{\rm gg}^2+\sigma_{\rm qg}(s)\mu_{\rm qg}^2
+\sigma_{\rm qq}(s)\mu_{\rm qq}^2\right]. 
\ee
 Hence, from ref. \cite{BK}, we have the three factorization theorems
\begin{eqnarray}
\frac{\signn(s)}{\siggp(s)}&=&\frac{\siggp(s)}{\siggg(s)}\label{eq:signn}\\
\frac{\Bnn(s)}{\Bgp(s)}&=&\frac{\Bgp(s)}{\Bgg(s)}\label{eq:Bnn}\\
\frac{\rhonn(s)}{\rhogp(s)}&=&\frac{\rhogp(s)}{\rhogg(s)}\label{eq:rhonn},
\end{eqnarray}
valid for {\em all} $s$.
It is easily  inferred from ref. \cite{BK} that 
\begin{eqnarray}
\siggg(s)&=\kappa P_{\rm had}^\gamma\siggp(s)&={\left(\kappa P_{\rm had}^\gamma\right)}^2\signn(s)\label{eq:sigfact}\\
\Bgg(s)&=\kappa\Bgp(s)&=\kappa^2\Bnn(s)\label{eq:Bfact}\\
\rhogg(s)&=\rhogp(s)&=\rhonn(s)\label{eq:rhofact},
\end{eqnarray}
where $P_{\rm had}^\gamma$ is the probability that a photon transforms into a hadron, assumed to be independent of energy and $\kappa$ is a proportionality constant, also independent of energy.  The value of $\kappa$, of course, is model-dependent.   For the case of the additive quark model, $\kappa=\frac{2}{3}$.

We emphasize the importance of the result of \eq{eq:rhofact} that the $\rho$'s are all equal,  independent of the assumed value of $\kappa$, {\em i.e.,} the equality does {\em not} depend on the assumed model.   
\subsection{Experimental Verification of Factorization using Compton Scattering}

\begin{figure}[htbp] 
\begin{center}
\mbox{\epsfig{file=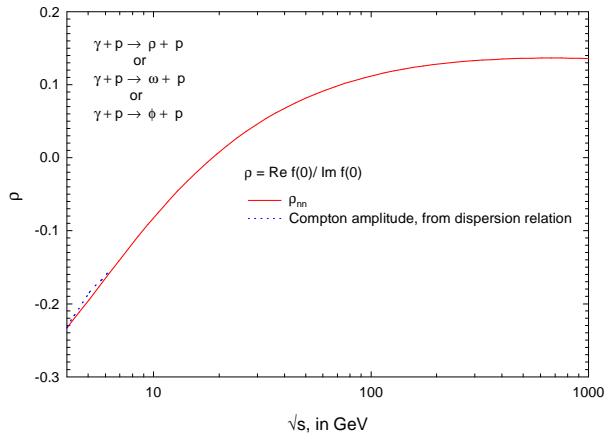,width=3.4in,%
bbllx=70pt,bblly=240pt,bburx=520pt,bbury=550pt,clip=}}
\end{center}
\caption[] {\footnotesize The solid curve is $\rho_{\rm nn}$, the predicted ratio of the 
real to imaginary part of the forward scattering amplitude for the
`elastic' reactions , $\gamma +p\rightarrow V + p$ scattering
amplitude, where $V$ is $\rho$, $\omega$ or $\phi$ (using the factorization theorem of \eq{eq:rhofact}). 
The dotted curve is ratio of the real to imaginary
part of the forward scattering amplitude for Compton
scattering , $\gamma +p\rightarrow\gamma + p$,
found from dispersion relations{\protect{\cite{gilman}}}.  It has been slightly displaced from
the solid curve for clarity in viewing.}
\label{fig:rhocompton}
\end{figure}

The solid
curve in \fig{fig:rhocompton} %
is $\rhonn$,   plotted as a function of the c.m.s. energy $\sqrt s$. According to \eq{eq:rhofact}, this should be the same as $\rhogp$. No experimental data for the `elastic scattering' reactions $\gamma +p\rightarrow V +p$, where $V$ is the vector meson $\rho$, $\omega$ or $\phi$, are available for direct comparison.  However, Damashek
and Gilman \cite{gilman} have calculated the $\rho$ value for Compton
scattering  $\gamma +p\rightarrow\gamma +p$ using dispersion relations, {\em i.e.,} the
{\em true} elastic scattering reaction for photon-proton scattering. The dispersion relation calculation gives $\rhogp$ if we assume that it is the same as that for the `elastic scattering' reactions $\gamma +p\rightarrow V +p$. In this picture we expect that $\Bgp=B_\rho=B_\omega=B_\phi$.
We then compare the dispersion relation calculation, the dotted line in \fig{fig:rhocompton},
with our prediction for  $\rhogp$ from \eq{eq:rhofact} ($\rhonn$, the solid line taken from ref. \cite{blockcr}).  The agreement is so
close that the two curves had to be  moved apart so that they may be
viewed more clearly.  It is clearly of importance to extend the energy region of the dispersion calculation. However, over the limited energy range available from the dispersion calculation, the prediction from \eq{eq:rhofact} of equal $\rho$-values is well verified experimentally. 
\subsection{Quark Counting}
\begin{figure}[htbp] 
\begin{center}
\mbox{\epsfig{file=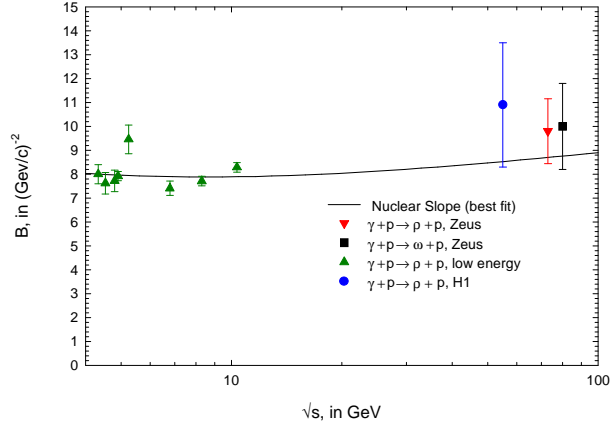,width=3.4in,%
bbllx=62pt,bblly=250pt,bburx=520pt,bbury=570pt,clip=}}
\end{center}
\caption {\footnotesize A $\chi^2$ fit of experimental data for  the nuclear slopes $B$, from the `elastic scattering' reactions $\gamma +p\rightarrow V + p$,
where $V$ is $\rho$, $\omega$ or $\phi$, to the relation $\Bgp=\kappa \Bnn$, of \eq{eq:Bfact}, where $\kappa=0.661\pm 0.008$ }
\label{fig:Bgp}
\end{figure}

The additive quark model tells us from quark counting  that $\kappa$ in \eq{eq:Bfact} is given by $\kappa=\frac{2}{3}$. We can experimentally determine $\kappa$ by invoking from \eq{eq:Bfact} the relation $\Bgp=\kappa\Bnn$ ($\Bnn$ is computed using the parameters from ref. \cite{blockcr}), and fitting $\kappa$.  In our picture, the `elastic scattering' reactions $\gamma +p\rightarrow V +p$, where $V$ is the vector meson $\rho$, $\omega$ or $\phi$, require that  $B_\rho=B_\omega=B_\phi(=\Bgp)$.  To determine the value of $\kappa$ in the relation $\Bgp=\kappa\,\Bnn$, a $\chi^2$ fit was made to the available $\Bgp$ data.  In Fig. \ref{fig:Bgp} we plot  $\kappa \Bnn$ {\em vs.} the c.m.s. energy $\sqrt s$, using  the best-fit value of  $\kappa=0.661$, against the experimental values of $\Bgp$.  The fit gave $\kappa=0.661\pm 0.008$, with a total $\chi^2=16.4$ for 10 degrees of freedom.  Inspection of Fig. \ref{fig:Bgp} shows that the experimental point of $B_\rho$ at $\sqrt s=5.2$ GeV--- which contributes 6.44 to the $\chi^2$---clearly cannot lie on any smooth curve and thus can  safely be ignored.  Neglecting the contribution of this point gives a $\chi^2$/d.f.=0.999, a very satisfactory result.  We emphasize that the experimental $\gamma p$ data thus
\begin{itemize}
\item  require  $\kappa=0.661\pm 0.008$, a $\approx 1\%$ measurement in excellent agreement with the value of 2/3 that is obtained from the additive quark model.
\item clearly verify the nuclear slope factorization theorem of \eq{eq:Bfact} over the available energy range spanned by the data.
\end{itemize}

For additional evidence involving the equality of the nuclear slopes $B_\rho$, $B_\gamma$ and $B_\phi$ from  differential elastic scatttering  data $\frac{d\sigma}{dt}$, see Figures (13,14,15) of ref. \cite{blockhalzenpancheri}.
\subsection{Vector Dominance, using $\gamma p$ Cross Sections}   
\begin{figure}[htbp] 
\begin{center}
\mbox{\epsfig{file=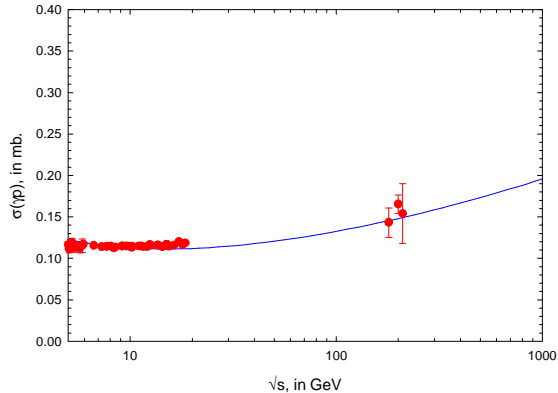,width=3.4in,%
bbllx=50pt,bblly=240pt,bburx=545pt,bbury=570pt,clip=}}
\end{center}
\caption {\footnotesize The total cross section for $\gamma p$ scattering. 
The solid curve is the predicted total cross section from the factorization relation of \eq{eq:sigfact}, $\siggp=\frac{2}{3}P_{\rm had}^{\gamma}\,\signn$, where $P_{\rm had}^\gamma=1/240$.}
\label{fig:2sigtot}
\end{figure}

Using $\kappa=\frac{2}{3}$ and \eq{eq:sigfact}, we write
\begin{equation}
\siggp(s)=
\frac{2}{3}P_{\rm had}^\gamma \,\signn(s),
\label{eq:sigtotgammap}
\end{equation}
where $P_{\rm had}^\gamma$ is the probability that a photon will interact as
a hadron. We will use the value $P_{\rm had}^\gamma=1/240$.  This value is
approximately 4\% greater than that derived from vector dominance, 1/249. Using (see
Table XXXV, pag.\ 393 of ref.\ \cite{bauer}) $f_{\rho}^2/4\pi=2.2$,
$f_{\omega}^2/4\pi=23.6$ and $f_{\phi}^2/4\pi=18.4$,
we find $\Sigma_{V}(4\pi\alpha/f_V^2)=1/249$, where
$V=\rho,\omega,\phi$.  The value we use of 1/240 is found by
normalizing the total $\gamma p$ cross section to the low energy data and is illustrated in Fig. \ref{fig:2sigtot}, where we plot the total cross section for $\gamma +p\rightarrow {\rm hadrons}$ from \eq{eq:sigtotgammap} as a function of the c.m.s. energy $\sqrt s$. The values for $\signn$ have been deduced from the results of ref. \cite{blockcr}, using the even eikonal from \eq{eq:chieven}.
The fit is exceptionally good, reproducing
the rising cross section for $\gamma p$, using the parameters fixed by
nucleon-nucleon scattering. The fact that we use the value 1/240 rather than 1/249 (4\% greater than the vector meson prediction) reflects the fact that 
$P_{\rm had}^\gamma$, the {\em total} probability that the photon is a hadron, should have a small contribution from the continuum, as well as from the vector mesons $\rho$, $\phi$ and $\omega$.  Thus, within the uncertainties of our calculation, the experimental data in the $\gamma p$ sector
\begin{itemize}
\item are compatible with vector meson dominance.
 \item agree with  cross section factorization theorem of \eq{eq:sigfact}.
\end{itemize}
%
%
\subsection{Experimental Verification of Factorization using $\gamma\gamma$ Scattering}
\begin{figure}[h]
\begin{center}
\mbox{\epsfig{file=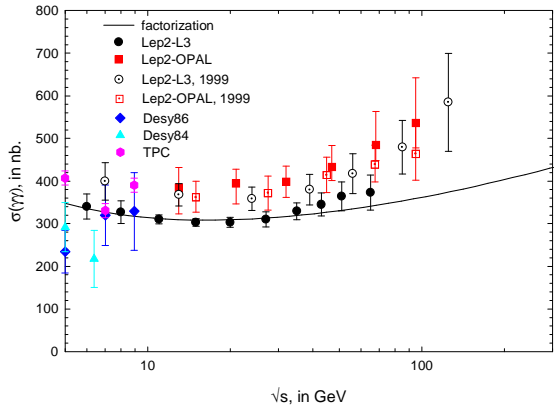,width=3.4in,%
bbllx=80pt,bblly=325pt,bburx=535pt,bbury=650pt,clip=}}
\end{center}
\caption {\footnotesize The predicted total cross section for $\gamma
\gamma$ scattering from the factorization theorem of \eq{eq:sigfact}, $\siggg={\left(\frac{2}{3}P_{\rm had}^\gamma\right)}^2\, \signn$, where $P_{\rm had}^\gamma=1/240$. The data sources are indicated in the legend.}
\label{fig:3sigtot}
\end{figure}

Using quark counting and the factorization theorem of \eq{eq:sigfact}, we now write
%
$\siggg={\left(\frac{2}{3}P_{\rm had}^\gamma\right)}^2\, \signn$
where $P_{\rm had}^\gamma=1/240$. In \fig{fig:3sigtot} %
we plot our factorization prediction for
$\siggg(s)$ as a function of the c.m.s.
energy $\sqrt s$ and compare it to  various sets of
experimental data. It is clear that factorization, as expressed in \eq{eq:sigfact}, selects the preliminary L3 data (solid circles) rather than the preliminary OPAL results (solid squares) \cite{exp}. The Monte Carlo-averaged final results of L3, given by the open circles, agrees, within errors, with the revised  OPAL data, with both new sets having a normalization of about 15-20 \% higher than the factorization prediction given by the solid line. The major difference between the earlier L3 result and the revised data was the use of the {\em average} normalization from the output of two {\em different} Monte Carlos. We find it  remarkable that the cross section factorization theorem of \eq{eq:sigfact}, using only input from the additive quark model and vector meson dominance, gives a reasonable prediction of the experimental data over a cross section magnitude span of more than a factor of $10^5$ and an energy region of $3\le\sqrt s\le 100$ GeV. On the other hand, a literal interpretation of the experimental data at the higher energies might indicate that the $\gamma\gamma$ cross section is rising slightly more rapidly than our prediction, a consequence perhaps of hard processes not accounted for by the vector dominance model. More accurate data are required for confirmation of this hypothesis. 
\section{Conclusions}
The available data on nn, $\gamma p$ and $\gamma\gamma$ reactions lend strong experimental support to the factorization hypotheses of 
\begin{itemize}
\item the well-known cross section factorization theorem of \eq{eq:signn} and \eq{eq:sigfact},
\item the lessor-known nuclear slope factorization theorem of \eq{eq:Bnn} and \eq{eq:Bfact}, 
\item the relatively obscure requirement of \eq{eq:rhofact} that $\rhonn=\rhogp=\rhogg$,
\end{itemize}
as well as 
\begin{itemize}
\item verifying the additive quark model by measuring $\kappa=0.661\pm 0.008$, a result within $\approx 1\%$ of the value of 2/3 expected for the quark model, using $\Bgp$ measurements over a wide span of energies, $3\le\sqrt s\le 200$.
\item confirming vector dominance using $\signn$, $\siggp$ and $\siggg$ over an energy region of $3\le\sqrt s\le 100$ GeV and a cross section factor of over $10^5$.

The QCD-inspired model\cite{blockhalzenpancheri} that we use fits the $pp$ and $\pbar p$ data on total cross sections, $\rho$-values and $B$ quite well and thus gives a good phenomological fit to those data.  We emphasize that the conclusions on factorization that we presented above are rooted in the available high energy {\em experimental} data for nn, $\gamma p$ and $\gamma\gamma$ collisions and do not depend on the details of the model used to fit nn data.  
\end{itemize}
\appendix
\section{Appendix}
\label{app:QCDeikonal}
For completeness, we summarize here the formulae and parameters needed to calculate nucleon-nucleon scattering, taken from ref. \cite{blockhalzenpancheri} and ref. \cite{blockcr}, which should be consulted for more detail.

We model the gluon-gluon contribution to the nucleon-nucleon cross section
following the parton model
\begin{equation}
\sigma_{gg}(s)=C_{gg}\int\Sigma_{gg}
\,\Theta(\hat s-m_0^2)F_{gg}(x_1,x_2)\,d\tau \; ,
\label{sigggqcd}
\end{equation}
where $\Sigma_{gg}=9\pi \alpha_s^2/m_0^2$ is a normalization constant,
${\hat s}=\tau s$, and
\begin{equation}
F_{gg}=\int\!\!\int f_{g}(x_1)f_{g}(x_2)
\delta (\tau - x_1 x_2)\,dx_1\,dx_2 \; .
\end{equation}
Note that for the parameterization of the gluon structure function as
$f_{g}(x)=N_g(1-x)^5/x^{1+\epsilon}$, where
$N_{g}=\frac{1}{2} (6-\epsilon)(5-\epsilon)\cdots(1-\epsilon)/5!$,
we can carry out explicitly the integrations, obtaining
\begin{equation}
\sigma_{\rm gg}(s)=C'_{\rm gg}\Sigma_{\rm gg}N_{\rm g}^2\sum_{i=0}^{5}
\left\{
\frac{a(i)-\frac{b(i)}{i-\epsilon}}{i-\epsilon}-\tau_0^{i-\epsilon}
\left(\frac{a(i)-\frac{b(i)}{i-\epsilon}}{i-\epsilon}
+\frac{b(i)}{i-\epsilon}\log (\tau_0)\right)
\right\}
\label{Fggintegrated}
\end{equation}
where $C'_{\rm gg}=C_{\rm gg}/9$, $\tau_0=m_0^2/s$,
$a(0)=-a(5)=-411/10$, $a(1)=-a(4)=-975/2,$
$a(2)=-a(3)=-600$, $b(0)=b(5)=-9,$ $b(1)=b(4)=-225,$ and
$b(2)=b(3)=-900$. The normalization constant $C'_{gg}$ is a fitted parameter
and the threshold mass $m_0$ is determined by experiment. The role of $m_0$, which is the onset of $\sigma_{gg}(s)$ with $\hat s$, is somewhat analagous to the role played by $p_{T}^{min}$ in the minijet models.  However, numerical exercises show that the value of $m_0$ is {\em not} dependent on energy and that the fit is not very sensitive to the value.  

Also our quark-quark and quark-gluon cross sections will be
parameterized following the scaling parton model. We approximate the quark-quark contribution with
\begin{equation}
\sigma_{qq}(s)=\Sigma_{gg}
\left( C + C_{Regge}^{even} \frac {m_0}{\sqrt s}\right) \; ,
\label{sigmaqq}
\end{equation}
where $C$ and $C_{Regge}^{even}$ are parameters to be fitted.
The quark-gluon interaction is approximated
as
\begin{equation}
\sigma_{qg}(s)=\Sigma_{gg}  C_{qg}^{log}\log\frac{s}{s_0},
\label{sigmaqg}
\end{equation}
where the normalization constant $C_{qg}^{log}$ and the square of the
energy scale in the log term $s_0$ are parameters are parameters to be fitted.

In summary, the even contribution to the eikonal is
\begin{eqnarray}
\chi_{even} &=&  i \left\{
\sigma_{gg}(s) W(b\,;\mu_{gg})
+\Sigma_{gg}\left( C+C_{Regge}^{even}\frac{m_0}{\sqrt{s}}\right)
W(b\,;\mu_{qq})
\right. \nonumber \\ && \left.
+\Sigma_{gg} C_{qg}^{log}
\log\frac{s}{s_0}W(b\,;\sqrt{\mu_{qq}\mu_{gg}})\right\} \; .
\label{finaleven}
\end{eqnarray}
The total even contribution is not yet analytic.
For large $s$, the {\rm even} amplitude in \eq{eq:chieven} is made
analytic by the substitution (see  the table on p. 580 of reference
\cite{bc}, along with reference \cite{eden}) $s\rightarrow
se^{-i\pi/2}.$ The quark contribution $\chi_{\rm qq}(s,b)$ accounts for  
the constant cross section and a Regge descending component ($\propto  
1/\sqrt s$), whereas the mixed quark-gluon term $\chi_{\rm qg}(s,b)$
simulates diffraction ($\propto \log s$).  The gluon-gluon term
$\chi_{\rm gg}(s,b)$, which eventually rises as a power law
$s^\epsilon$,  accounts for the rising cross section and dominates at  
the highest energies. In \eq{eq:chieven}, the inverse sizes (in impact  
parameter space) $\mu_{\rm qq}$ and $\mu_{\rm gg}$ are determined by
experiment, whereas the quark-gluon inverse size is taken as
$\sqrt{\mu_{\rm qq}\mu_{\rm gg}}$.


\begin{table}[h]
\begin{tabular}{ll}
\hline
 Fitted              				& Fixed     \\ \hline
$C=5.65\pm 0.14 $       			& $\alpha_s=0.5$   \\
$C^{log}_{qg}=0.103 \pm 0.026$      	& $\epsilon=0.05$  \\
$C'_{gg}=(1.12 \pm 0.05) \times 10^{-3}$ 	& $m_0=0.6$ GeV\\
$C^{even}_{Regge}=25.3 \pm 2.0$		& $\mu_{qq}=0.89$ GeV  \\
$C_{odd}=7.62 \pm 0.28 $ 			& $\mu_{gg}=0.73$ GeV  \\
$s_0=16.9\pm 4.9$ GeV$^2$			&$\mu_{odd}=0.53$ GeV \\
							&$k=1.349\pm0.045$\\
\hline
\end{tabular}
\caption{Value of the parameters.}
\label{tab:param}
\end{table}
In addition to the parameters in ref. \cite{blockhalzenpancheri}, the cosmic ray data fit requires the specification of another parameter  $k$,  the proportionality constant between the measured mean free path (in air) and the true mean free path in air.  The major difference between the parameters of Table \ref{tab:param} and the parameters of ref. \cite{blockhalzenpancheri} is that the errors of $C'_{gg}$ and $s_0$ are now smaller by a factor of $\approx 2$, due to the large lever arm of the hign energy cosmic ray points.  This in turn  leads to significantly smaller errors in our predictions for high energy cross sections, since the dominant term at high energies is $\sigma_{\rm gg}(s)$.

\newpage

%

%
\end{document}